\documentclass[11pt,a4paper]{article}

\usepackage{fontspec}
\usepackage{amsmath,amssymb}
\usepackage{booktabs}
\usepackage{array}
\usepackage{enumitem}
\usepackage{float}
\usepackage{graphicx}
\usepackage{microtype}
\usepackage[numbers,sort&compress]{natbib}
\usepackage{xcolor}
\usepackage{xurl}
\usepackage{hyperref}

\hypersetup{
  colorlinks=true,
  linkcolor=blue!55!black,
  citecolor=blue!55!black,
  urlcolor=blue!55!black,
  pdftitle={When Should Multi-Round RAG Stop? Structured Stopping Judgments and Retrieval Reduction in Search-R1}
}

\newcommand{\model}[1]{\texttt{#1}}

\title{When Should Multi-Round RAG Stop?\\Structured Stopping Judgments and Retrieval Reduction in Search-R1}
\author{Weimeng Luo\\
\small Unaffiliated\\
\small \href{mailto:weimengluo@gmail.com}{weimengluo@gmail.com}}
\date{}

\begin{document}
\maketitle

\begin{abstract}
Multi-round retrieval-augmented generation (RAG) must decide when to stop searching as evidence accumulates. Because the deployed policy is determined by the first STOP on each trajectory, this is a sequential selection problem rather than an independent state-classification task. We adapt S2G-RAG's structured sufficiency-and-gap judgment to a frozen Search-R1 pipeline and train a Qwen3.5-2B judge on 3,009 states from 900 disjoint HotpotQA questions. Search-R1's reasoner, retriever, corpus, prompt, and search budget remain unchanged, while the judge checkpoint and stopping threshold are selected on grouped validation and frozen before confirmatory evaluation. On the confirmatory test set, the resulting policy reduces retrieval calls by 77 (3.70\%) relative to Native Search-R1, while Official Exact Match decreases by 0.625 percentage points. Thus, the trained S2G-style structured judge reduces retrieval while broadly preserving answer accuracy. The result does not imply unchanged or improved accuracy, safe stopping, or lower total inference cost.
\end{abstract}


\section{Introduction}

Iterative retrieval lets a language model decide which evidence to request as reasoning unfolds. IRCoT interleaves retrieval and reasoning \citep{trivedi2023ircot}. Self-RAG learns retrieval and reflection decisions \citep{asai2023selfrag}, Adaptive-RAG routes questions among strategies of different complexity \citep{jeong2024adaptiverag}, and Search-R1 learns to reason and call a search engine through reinforcement learning \citep{jin2025searchr1}. These methods improve flexibility, but they also create a stopping problem. Retrieving too little can leave the answer unsupported. Retrieving after a sufficient state consumes latency and compute, and the additional context can change a correct candidate into an incorrect one.

Recent work has made the stopping decision explicit. SIM-RAG learns whether to continue search from generated answers and rationales \citep{yang2025simrag}. S2G-RAG asks a judge to predict both sufficiency and the missing information needed for the next retrieval \citep{li2026s2grag}. TASR studies training-free stopping for iterative retrieval \citep{kieback2026tasr}, while RAG-Critic uses critic feedback to guide an agentic workflow \citep{dong2025ragcritic}. These systems motivate learned stopping, but their reported state-level scores or end-to-end averages do not by themselves identify why a deployed stopping policy succeeds or fails.

The missing link is sequential selection. A state classifier is evaluated over many intermediate states, whereas an online policy is determined by the first threshold crossing on each trajectory. One early false positive can suppress every later state. Conversely, a perfectly conservative classifier can leave the system unchanged. The feasibility of improvement is also bounded by candidate reachability: if no correct answer can be generated from any frozen reachable state, changing the stopping time cannot make the question correct. This paper therefore asks four questions:

\begin{enumerate}[leftmargin=*]
  \item How often does an audited strong search policy continue after a correct answer is already reachable?
  \item How much quality and retrieval-cost headroom exists when only the stopping time may change?
  \item Does structured sufficiency-and-gap supervision improve state ranking and the deployed earliest-stop policy?
  \item What answer risk accompanies the retrieval reduction that survives confirmatory evaluation?
\end{enumerate}

We answer these questions in two stages. First, an exploratory reachability audit runs a frozen, non-intervening answer-only probe at native Search-R1 states. It estimates empirical conditional oracles without changing the search trajectory. Second, a scale-up experiment trains an S2G judge on disjoint training questions, freezes model and threshold choices before final evaluation, and reserves indices 200--999 of a 1,000-question HotpotQA set for confirmatory analysis.

The paper makes three contributions. First, it adapts an S2G-style structured judging mechanism to Search-R1 and trains a Qwen3.5-2B judge specifically on Search-R1 states. Second, it provides a trajectory-aware formulation that separates candidate reachability, state ranking, first-crossing policy behavior, answer quality, and cost, including a controlled mechanism ladder in which binary supervision improves average precision but harms the system. Third, it shows on the confirmatory test set that the complete policy reduces retrieval calls while keeping the loss in answer accuracy within the prespecified two-percentage-point tolerance. Safe stopping is not established.

\section{Related Work and Positioning}

\subsection{Adaptive and Iterative Retrieval}

Adaptive retrieval encompasses several decision interfaces. Query-level routers decide among no retrieval, one retrieval, and multi-step retrieval before the trajectory starts \citep{jeong2024adaptiverag}. Interleaved systems decide what to retrieve during reasoning \citep{trivedi2023ircot,jin2025searchr1}. Generation-time systems emit control or reflection tokens that can trigger retrieval or critique \citep{asai2023selfrag}. Answer-level controllers instead decide whether an existing candidate should be accepted or whether search should continue \citep{yang2025simrag}. These interfaces are related, but their oracles and error semantics differ. An answer-level STOP label cannot be transferred to a pre-query router without changing the decision problem.

Search-R1 is the upstream policy studied here \citep{jin2025searchr1}. Our objective is not to improve its reasoner or retriever. We freeze the 7B reasoner, E5 top-3 retrieval, Wiki-2018 corpus, prompting, and four-search budget. The only intervention is whether a reachable state is accepted early. This restriction makes the resulting oracle conditional on the frozen interface rather than a global upper bound over possible queries, documents, or models.

\subsection{Learned Judges and Structured Gaps}

SIM-RAG directly models whether a search process should continue \citep{yang2025simrag}. Our released-checkpoint pilot treats it as an external answer-level controller and finds a cross-system transfer failure: on 200 questions, it performs eight harmful changes and no rescues relative to native Search-R1. We retain that result only as exploratory evidence because its trajectories, reasoner, and interface differ from the training environment of the released critic.

S2G-RAG argues that a binary sufficient/insufficient target discards information about what remains missing \citep{li2026s2grag}. We transfer this structured judging mechanism to the Search-R1 state interface and train a judge specifically on Search-R1 states. Our judge outputs a frozen JSON schema with a Boolean \model{sufficient} field and a list of \model{gap\_items}. The online policy uses the log-probability margin for the Boolean field at a frozen prefix, while the structured target regularizes training toward explicit missing-information judgments. We do not claim to reproduce the complete S2G-RAG system, nor do we train or modify Search-R1 itself; the evaluated system is the complete policy formed by frozen Search-R1 and the trained S2G-style judge.

\subsection{Risk, Coverage, and System Utility}

Selective classification separates the conditional error of accepted predictions from their coverage \citep{geifman2017selective}. Certified risk work extends this concern to RAG generation \citep{kang2024crag}. We use the same conceptual separation, but do not claim a formal risk guarantee. The confirmatory gate controls an EM non-inferiority margin and tests whether paired search calls decrease. STOP precision, early-stop safety, average precision, calibration, and end-to-end EM remain distinct outcomes.

This positioning matters because a retrieval reduction is not equivalent to total efficiency. The judge itself consumes inference compute, and our experiment does not normalize wall-clock time, energy, or token-equivalent judge cost. The paper therefore reports retrieval calls as the cost endpoint and reserves ``total efficiency'' for future matched-compute evaluation.

\section{Problem Formulation}

\begin{figure}[t]
\centering
\includegraphics[width=\linewidth]{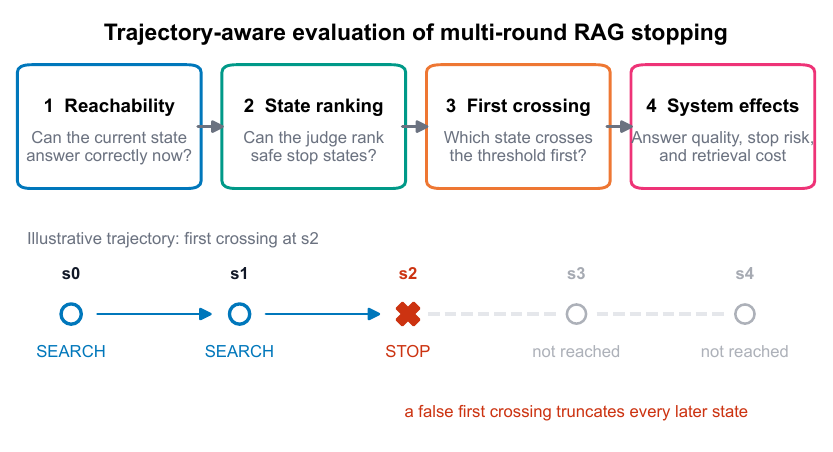}
\caption{Trajectory-aware evaluation of multi-round RAG stopping. A correct
reachable candidate is necessary but not sufficient; the judge must rank it,
the earliest threshold crossing must select it, and the resulting answer quality,
selected-stop risk, and retrieval cost must be evaluated separately.}
\label{fig:stopping-framework}
\end{figure}

Figure~\ref{fig:stopping-framework} summarizes the evaluation chain. Each layer
answers a different question: reachable candidates bound what stopping can
achieve, state scores diagnose representation quality, the first threshold
crossing determines the deployed decision, and system outcomes establish the
quality--risk--cost trade-off.

\subsection{Reachable States and Stop-Now Answers}

For question $q_i$, native Search-R1 produces a trajectory of reachable states
\begin{equation}
  z_{i,t}=(q_i,C_{i,t},h_{i,t},a^{\mathrm{native}}_{i,t}),
\end{equation}
where $C_{i,t}$ is the accumulated retrieved context, $h_{i,t}$ is the native reasoning history, and $a^{\mathrm{native}}_{i,t}$ is the next native action. We consider the initial state and each state after a successful retrieval. A frozen answer-only probe $g$ maps the question and current context to a candidate answer $\tilde a_{i,t}=g(q_i,C_{i,t})$. The probe output never re-enters the native trajectory.

Let $y_{i,t}=1$ when $\tilde a_{i,t}$ matches the official answer under benchmark normalization. A safe early-stop opportunity exists when native Search-R1 proposes another SEARCH while $y_{i,t}=1$. This target does not assert that the retrieved evidence is complete in a supporting-fact sense. It states only that the frozen answerer can already recover the official answer from the current interface.

\subsection{Conditional Oracles}

The quality oracle selects the earliest reachable correct probe answer, if one exists. Its EM is an empirical upper bound for answer selection over the frozen states. The native-preserving cost oracle applies an earlier correct stop only to questions that native Search-R1 already answers correctly. It therefore holds native EM fixed and estimates how many search calls can be removed without changing those final labels.

Neither oracle is deployable. Both use labels after trajectory generation. Their role is diagnostic: when quality and cost headroom are negligible, a learned stopping model has little room to improve the frozen interface.

\subsection{Scores, Earliest Stopping, and Metrics}

The judge produces a scalar score $s_{i,t}$. For threshold $\theta$, the online policy stops at
\begin{equation}
  \tau_i(\theta)=\min\{t:s_{i,t}\geq\theta\}.
\end{equation}
If no state crosses the threshold, execution follows the frozen native fallback. State-level average precision (AP) evaluates ranking over eligible states. STOP precision and recall evaluate the thresholded state decision. System-level early-stop safety counts whether the selected first crossing is a safe opportunity. End-to-end quality is official EM over all questions, and cost is the mean number of search calls.

The main paired effects for model $m$ relative to native execution are
\begin{align}
  \Delta_{\mathrm{EM}}(m)&=\frac{1}{N}\sum_i\left[\mathrm{EM}_{i,m}-\mathrm{EM}_{i,\mathrm{native}}\right],\\
  \Delta_{\mathrm{search}}(m)&=\frac{1}{N}\sum_i\left[S_{i,m}-S_{i,\mathrm{native}}\right].
\end{align}
Confidence intervals use 10,000 question-level paired bootstrap resamples with seed 20260728.

\section{Experimental Design}

\subsection{Audited Upstream System}

The upstream agent is the official Search-R1 Qwen2.5-7B PPO v0.2 checkpoint with E5 top-3 retrieval over Wiki-2018 and at most four searches \citep{jin2025searchr1}. A separate 1,200-question NQ/HotpotQA audit obtained official EM of 45.67\% and 46.00\%, respectively. The reported reference values fell inside source-specific Wilson 95\% intervals and differed by less than five percentage points. We call this status \emph{configuration-verified and compatible}, not a strict reproduction.

The main scale-up data use the first 1,000 questions of the HotpotQA distractor development set \citep{yang2018hotpotqa}, materialized through the S2G-code-aligned preprocessing path. Question identity and normalized text were checked against every training, grouped-validation, and reserve question.

\subsection{Exploratory Reachability Audit}

Gate H0 reuses 200 frozen native trajectories. It evaluates 723 native reachable states: 200 initial states and 523 states preceding native SEARCH actions. Seventy-five states created only by a forced-continue intervention are retained as a counterfactual supplement and excluded from native-behavior attribution. Probe prompting, deterministic decoding, parsing, state deduplication, and cost accounting were frozen before labels were made available. All 798 probes parsed successfully.

H0 is exploratory because its 200 labels had already been used for earlier pilot analyses. It determines whether further stopping research is warranted and supplies failure decompositions, but it does not support the final system claim.

\subsection{Training the Expanded Structured Judge}

The training source is the HotpotQA training split, separated by question into 700 fresh training questions, 100 grouped-validation questions, and 200 reserve questions. The reserve was activated according to a predeclared data-sufficiency gate, yielding 900 training questions and 3,009 clean states. Teacher outputs with generation or schema failures were removed by frozen rules. Labels were not manually repaired. Supporting-fact labels were used only for one-way conflict filtering after teacher generation and were never provided to the judge.

The judge is Qwen3.5-2B. It is trained for three epochs (12,654 steps) to emit the top-level key order \model{sufficient} followed by \model{gap\_items}. All losses are finite and no training OOM occurs. Epoch 1 is selected using grouped teacher-target token negative log-likelihood, without using answer labels or final-set metrics. The frozen expanded adapter tree has SHA-256 \nolinkurl{cfed584b95025cc4f1fd555cd2236b2773f1289c67dab03d26b2829659578767}.

We compare three judges: the unadapted Structured Base model, an older S2G LoRA retained only as a historical comparator, and the expanded S2G LoRA. Each receives the same state representation and score extraction. No final labels are used to alter prompts, checkpoints, thresholds, budgets, input order, or statistics.

\subsection{Threshold Freezing and Confirmatory Analysis}

For each judge, grouped validation chooses the threshold that maximizes recall subject to empirical STOP precision of at least 0.90 and at least 10 predicted STOP states. If no threshold is feasible, the frozen implementation encodes \model{fallback=never\_stop} as a finite sentinel equal to the maximum validation score plus one. This guarantees zero STOP predictions on grouped validation, but not literal never-stopping under distribution shift. The expanded judge freezes at a margin of approximately 7.875. Structured Base and the old LoRA use sentinels of 2.5 and approximately 8.0, respectively.

The first 200 development questions remain exploratory. Indices 200--999 form the confirmatory test set. Before evaluation, we specify that the expanded policy may lose at most two EM percentage points relative to Native. Formally, the lower endpoint of the paired 95\% interval for $\Delta_{\mathrm{EM}}$ must exceed $-0.02$. The upper endpoint of the paired 95\% interval for $\Delta_{\mathrm{search}}$ must also be below zero. The first condition is the statistical non-inferiority criterion; in plain language, the loss in answer accuracy must remain within the predefined acceptable range. It does not require accuracy to be exactly unchanged. Full 1,000-question results are descriptive only.

Final inference follows a blind sequence. Native trajectories, answer-only probes, and all judge scores undergo structure and order audits before final labels are isolated for evaluation. A preblind script computed a SHA-256 hash of the final-label file despite the no-read rule. The hash was not used for selection and the file was not mounted into inference containers, but the access is a procedural nonconformance and is disclosed rather than silently waived.

\section{Results}

\subsection{A Strong Search Policy Still Has Stopping Headroom}

On the exploratory H0 set, 143 of 523 native SEARCH actions (27.3\%) occur at states where the frozen probe already produces the correct official answer. These opportunities occur in 70 of 200 questions. Sixty questions show cost-only over-search: an early probe and the native final answer are both correct. Ten questions show harmful over-search: an earlier probe is correct but the native final answer is wrong.

Native EM is 46.5\%. The quality oracle reaches 51.5\%, an exact gain of 10/200 or 5.0 percentage points. The native-preserving cost oracle removes 131 of 523 searches (25.05\%) while holding native EM fixed. Fixed budgets do not realize the same question-specific trade-off: answer-only EM at budgets $k=0,1,2,3,4$ is 18.5\%, 34.0\%, 43.0\%, 46.0\%, and 47.0\%. This gap between a per-question oracle and uniform budgets motivates a learned state-dependent controller.

\begin{figure}[t]
\centering
\includegraphics[width=\linewidth]{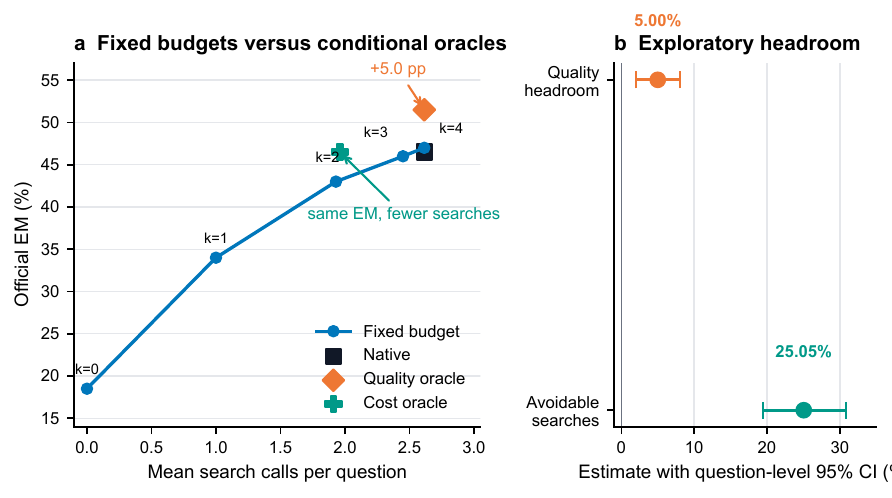}
\caption{Exploratory H0 stopping headroom on 200 questions. Panel (a) compares
uniform search budgets with native Search-R1 and two label-dependent conditional
oracles. Panel (b) reports question-level bootstrap 95\% intervals. The quality
oracle is retrospective and the cost oracle preserves Native EM; neither is a
deployable policy or a confirmatory result.}
\label{fig:h0-headroom}
\end{figure}

\subsection{State Ranking Does Not Determine the Policy}

The exploratory mechanism ladder exposes a ranking--policy mismatch. A binary LoRA raises STOP AP from 0.732 to 0.828 on the Core200 diagnostic, yet the zero-threshold policy reduces official EM from 36.5\% to 31.5\% and mean searches from 2.635 to 1.430. The model has learned a stronger state ranking signal but stops too aggressively under the deployed threshold.

Replacing the binary target with a minimal S2G target reduces wrong STOP decisions from 99 to 24 and premature STOP decisions from 18 to 4. Official EM recovers from 31.5\% to 35.5\%, while mean searches increase from 1.430 to 2.390. Relative to Structured Base, however, the EM difference is only $+0.015$ with a paired 95\% interval of $[-0.010,0.040]$. This diagnostic supports structured gaps as a mechanism for reducing over-stopping, not a system-level superiority claim.

\subsection{Grouped Validation Freezes a Conservative Policy}

On 100 grouped-validation questions, the expanded judge obtains STOP precision 0.9091, recall 0.1205, and AP 0.5285 at its frozen threshold. Its end-to-end EM equals Native Search-R1 at 0.53, and mean search calls fall from 2.45 to 2.33. Structured Base and the old LoRA cannot satisfy the validation precision-and-support rule and therefore receive the finite conservative sentinels described above. These measurements select the final policy. They are not part of the confirmatory claim.

\subsection{Confirmatory Test Set: Limited Accuracy Loss and Fewer Searches}

Both confirmatory criteria pass. On the confirmatory test set comprising indices 200--999, Native Search-R1 reaches EM 0.44875 with 2.60125 mean search calls, whereas the expanded S2G policy reaches EM 0.44250 with 2.50500 searches. The paired EM difference is $-0.00625$ (95\% CI $[-0.01250,0]$): the point estimate loses 0.625 percentage points, and the interval remains within the predeclared maximum acceptable loss of two percentage points. The search-call difference is $-0.09625$ (95\% CI $[-0.12000,-0.07375]$); its upper endpoint is below zero.

In absolute terms, Native makes 2,081 retrieval calls and the expanded policy makes 2,004, a reduction of 77 calls or 3.70\%. Six questions change from Native-correct to policy-wrong, while one changes from Native-wrong to policy-correct, yielding the net $-5/800=-0.625$ percentage-point EM difference. Table~\ref{tab:suffix800} places each effect, interval, and frozen criterion together so that non-inferiority is not read as equivalence or superiority.

\begin{table}[H]
\centering
\small
\caption{Primary results on the confirmatory test set. Differences are Expanded S2G minus Native; both frozen criteria pass.}
\label{tab:suffix800}
\resizebox{\linewidth}{!}{%
\begin{tabular}{lrrrll}
\toprule
Metric & Native & Expanded & Difference [95\% CI] & Frozen criterion & Verdict \\
\midrule
Official EM & 0.44875 & 0.44250 & $-0.00625\;[-0.01250,0]$ & CI lower $>-0.02$ & PASS \\
Mean searches & 2.60125 & 2.50500 & $-0.09625\;[-0.12000,-0.07375]$ & CI upper $<0$ & PASS \\
\bottomrule
\end{tabular}
}
\end{table}

Expanded S2G improves STOP AP over Structured Base by 0.03983 (95\% CI $[0.00571,0.07228]$), providing confirmatory evidence of better state ranking. Ranking improvement does not establish safe early stopping. Table~\ref{tab:suffix800-diagnostics} separates state-level diagnostics from the system-level first selected stop. One test-set state exceeds each finite fallback sentinel for Structured Base and the old S2G model, so these baselines implement validation-relative conservative fallbacks rather than mathematically infinite thresholds.

\begin{table}[H]
\centering
\small
\caption{Judge diagnostics on the confirmatory test set. STOP P/R/AP are state-level metrics over all eligible states; early-stop counts evaluate the first threshold crossing per question.}
\label{tab:suffix800-diagnostics}
\resizebox{\linewidth}{!}{%
\begin{tabular}{lrrrr}
\toprule
Judge & STOP P & STOP R & STOP AP & Early stops (safe/unsafe) \\
\midrule
Structured Base & 0.0000 & 0.0000 & 0.36375 & 1 (0/1) \\
Old S2G LoRA & 1.0000 & 0.0019 & 0.36596 & 1 (1/0) \\
Expanded S2G LoRA & 0.6216 & 0.0880 & 0.40358 & 69 (42/27) \\
\bottomrule
\end{tabular}
}
\end{table}

\subsection{The Retrieval Reduction Carries Material Stop Risk}

The expanded policy makes 69 system-level early stops, or 8.63\% of the 800 questions. Forty-two are safe under the frozen target and 27 are unsafe, so the selected early-stop unsafe fraction is 39.13\%. The 42 safe selections cover only 14.29\% of the 294 questions that contain a safe opportunity. State-level STOP precision is 0.6216 and recall is 0.0880. The distinction between 42/69 system outcomes and state-level precision is intentional: the former evaluates the first selected stop per question, whereas the latter evaluates all eligible states under the threshold.

The grouped-validation STOP precision of 0.9091 is based on only 11 predicted STOP states and falls to 0.6216 at the frozen threshold on the confirmatory test set. The validation precision constraint therefore does not transfer to the final distribution. Calibration is also weak. Expanded S2G has Brier score 0.2952 and ten-bin expected calibration error (ECE) 0.2911, compared with 0.1992 and 0.1179 for Structured Base. The expanded judge ranks safe opportunities better, but its raw confidence scale is less calibrated. At a post-hoc state-level operating point that reaches at least 0.90 precision, recall falls to 0.0057. These facts rule out describing the policy as risk-certified or broadly safe.

The 27 unsafe early stops do not map one-to-one to newly wrong final questions. Relative to Native, six questions become wrong and one becomes correct. The remaining unsafe stops occur where the Native outcome is already wrong or where an early state fails the frozen safe-opportunity definition even though the final answer remains correct. The EM non-inferiority test measures average quality over all questions, whereas selected-stop risk measures conditional risk among the policy's active acceptances. They answer different questions and must both be reported.

\begin{figure}[t]
\centering
\includegraphics[width=\linewidth]{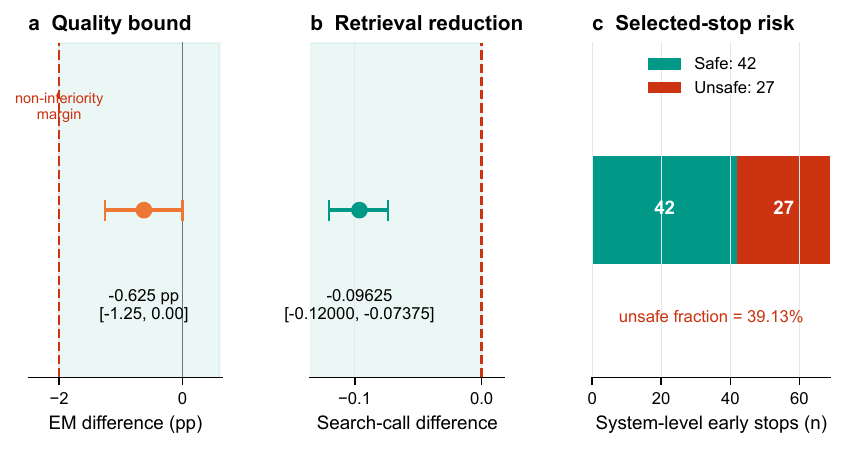}
\caption{Confirmatory test-set quality--cost--risk result for Expanded S2G
relative to Native Search-R1. Panels (a) and (b) show paired bootstrap 95\%
intervals; the dashed lines mark the frozen $-2$ percentage-point EM
non-inferiority margin and zero search difference. Panel (c) decomposes the 69
system-level first selected early stops. The result supports retrieval reduction
within the quality bound, not answer-quality superiority or safe stopping.}
\label{fig:confirmatory-tradeoff}
\end{figure}

Figure~\ref{fig:confirmatory-tradeoff} places these three facts together: the EM
interval remains above the frozen non-inferiority margin, the retrieval interval
excludes zero in the favorable direction, and the selected early stops retain a
material unsafe fraction. This is the paper's positive result and its principal
boundary in one view.

\subsection{Full 1,000-Question Results Are Descriptive}

Across indices 0--999, native EM/searches are 0.4490/2.6100 and expanded S2G obtains 0.4430/2.5210. Expanded STOP precision/recall/AP are 0.6163/0.0805/0.3987. The system makes 80 early stops, 49 safe and 31 unsafe. These values are useful for inventory and reproducibility, but the first 200 questions had exploratory status and cannot be pooled into the confirmatory claim.

\section{Discussion}

\subsection{What the Results Establish}

The experiments support a clear central claim. We adapt an S2G-style structured judging mechanism to frozen Search-R1 and train a Qwen3.5-2B judge on Search-R1 states. On the confirmatory test set, the resulting complete policy reduces retrieval calls while keeping the loss in answer accuracy within the two-percentage-point range specified before evaluation. The 95\% interval for the retrieval difference excludes zero, and the EM interval remains above the allowed-loss boundary. The reachability audit explains why Search-R1 has removable retrieval calls, while the ranking--policy diagnostics show why the judge must be evaluated through question-level first crossings. The system result belongs to the complete combination of training data, structured supervision, checkpoint, and threshold and cannot be attributed to any one component in isolation.

\subsection{What the Results Do Not Establish}

The expanded judge does not improve answer quality. Its point estimate is 0.625 percentage points below native EM. Keeping the loss within an acceptable range does not mean that accuracy is exactly unchanged, and it does not imply superiority.

The policy is also not a safe stopping rule. Twenty-seven of 69 early stops are unsafe under the frozen definition, and the 0.90 validation precision constraint does not transfer to final evaluation. A deployment that requires a low conditional error rate among accepted early stops would need a separate risk-control procedure, more calibration data, a more conservative policy, or abstention. The present quality gate can tolerate some unsafe STOP decisions because it averages over all questions.

Finally, fewer retrieval calls do not prove lower total cost. The Qwen3.5-2B judge runs on eligible states. Its compute, memory, and latency are not converted into retrieval-equivalent units. The result is best interpreted for environments where retrieval dominates marginal cost or where judge inference can be amortized. Matched-latency and matched-energy evaluation remain open.

\subsection{Implications for Stopping Research}

Future stopping work should report a four-part evaluation. First, establish candidate reachability and oracle headroom on the exact decision interface. Second, report state ranking to diagnose representation learning. Third, replay or execute the earliest-stop policy and report question-level safe and unsafe selections. Fourth, pair quality with retrieval, token, latency, and judge-compute costs. Any one of these views alone is incomplete.

Structured gap prediction remains promising because it reduces aggressive over-stopping in the mechanism ladder and improves AP in the scale-up study. Its next test should target the large gap between AP and calibrated decision utility. Candidate directions include calibration-aware training, trajectory-level losses that penalize the first unsafe crossing, and explicit cost-sensitive objectives. These are hypotheses for new experiments, not claims supported by the present study.

\section{Conclusion}

We adapt an S2G-style structured judging mechanism to frozen Search-R1 and train a Qwen3.5-2B judge specifically on Search-R1 states. On the confirmatory test set, the complete policy makes 77 fewer retrieval calls than Native Search-R1: mean searches fall from 2.60125 to 2.50500, a difference of $-0.09625$ or 3.70\% (95\% CI $[-0.12000,-0.07375]$). Official EM falls from 0.44875 to 0.44250, a loss of 0.625 percentage points (95\% CI $[-1.25,0]$ percentage points), which remains within the maximum acceptable loss of two percentage points specified before evaluation. The experiment therefore shows that the trained S2G-style judge reduces retrieval calls while broadly preserving answer accuracy. Here, ``broadly preserving'' allows a small loss within the prespecified range; it does not mean that accuracy is unchanged or improved.

The policy is not yet a safe stopping rule: 27 of its 69 early stops are unsafe, for selected-stop risk of 39.13\%. Keeping overall answer accuracy within an acceptable range and making every early stop safe are distinct objectives.

\section*{Ethics and Responsible Reporting}

The study uses public question-answering benchmarks and does not involve human participants or personal data. The main reporting risk is scientific overstatement. We therefore distinguish exploratory, validation, confirmatory, and descriptive populations, disclose infrastructure and blinding incidents, and state negative boundaries alongside positive results.

\section*{AI-Assistance Disclosure}

Generative AI tools were used for code assistance, experiment-monitoring support, language editing, and manuscript restructuring. The author retained responsibility for experimental design, execution authorization, evidence verification, statistical interpretation, and every scientific claim. AI-generated text was checked against the frozen result artifacts and cited primary sources.

\section*{Data, Code, Funding, and Competing Interests}

The public reproduction repository contains the S2G Judge/Search-R1 implementation, frozen protocols, aggregate results, and paper source: \url{https://github.com/luobostorm/search-r1-s2g-stopping}. Per-question trajectories, benchmark labels, and model weights are not redistributed. Model checkpoints and benchmark data remain subject to their original licenses. No external funding is declared. No competing interests are declared.

\bibliographystyle{unsrtnat}
\bibliography{references}

\begin{thebibliography}{11}
\providecommand{\natexlab}[1]{#1}
\providecommand{\url}[1]{\texttt{#1}}
\expandafter\ifx\csname urlstyle\endcsname\relax
  \providecommand{\doi}[1]{doi: #1}\else
  \providecommand{\doi}{doi: \begingroup \urlstyle{rm}\Url}\fi

\bibitem[Trivedi et~al.(2023)Trivedi, Balasubramanian, Khot, and
  Sabharwal]{trivedi2023ircot}
Harsh Trivedi, Niranjan Balasubramanian, Tushar Khot, and Ashish Sabharwal.
\newblock Interleaving retrieval with chain-of-thought reasoning for
  knowledge-intensive multi-step questions.
\newblock In \emph{Proceedings of the 61st Annual Meeting of the Association
  for Computational Linguistics (Volume 1: Long Papers)}, 2023.
\newblock \doi{10.18653/v1/2023.acl-long.557}.
\newblock URL \url{https://aclanthology.org/2023.acl-long.557/}.

\bibitem[Asai et~al.(2023)Asai, Wu, Wang, Sil, and Hajishirzi]{asai2023selfrag}
Akari Asai, Zeqiu Wu, Yizhong Wang, Avirup Sil, and Hannaneh Hajishirzi.
\newblock Self-{RAG}: Learning to retrieve, generate, and critique through
  self-reflection.
\newblock \emph{arXiv preprint arXiv:2310.11511}, 2023.
\newblock \doi{10.48550/arXiv.2310.11511}.
\newblock URL \url{https://arxiv.org/abs/2310.11511}.

\bibitem[Jeong et~al.(2024)Jeong, Baek, Cho, Hwang, and
  Park]{jeong2024adaptiverag}
Soyeong Jeong, Jinheon Baek, Sukmin Cho, Sung~Ju Hwang, and Jong Park.
\newblock Adaptive-{RAG}: Learning to adapt retrieval-augmented large language
  models through question complexity.
\newblock In \emph{Proceedings of the 2024 Conference of the North American
  Chapter of the Association for Computational Linguistics: Human Language
  Technologies (Volume 1: Long Papers)}, pages 7036--7050, 2024.
\newblock \doi{10.18653/v1/2024.naacl-long.389}.
\newblock URL \url{https://aclanthology.org/2024.naacl-long.389/}.

\bibitem[Jin et~al.(2025)Jin, Zeng, Yue, Wang, Zamani, and
  Han]{jin2025searchr1}
Bowen Jin, Hansi Zeng, Zhenrui Yue, Dong Wang, Hamed Zamani, and Jiawei Han.
\newblock {Search-R1}: Training {LLMs} to reason and leverage search engines
  with reinforcement learning.
\newblock \emph{arXiv preprint arXiv:2503.09516}, 2025.
\newblock URL \url{https://arxiv.org/abs/2503.09516}.

\bibitem[Yang et~al.(2025)Yang, Zeng, Rao, and Zhang]{yang2025simrag}
Diji Yang, Linda Zeng, Jinmeng Rao, and Yi~Zhang.
\newblock Knowing you don't know: Learning when to continue search in
  multi-round {RAG} through self-practicing.
\newblock \emph{arXiv preprint arXiv:2505.02811}, 2025.
\newblock URL \url{https://arxiv.org/abs/2505.02811}.

\bibitem[Li et~al.(2026)Li, Zou, Lv, Zhang, and Zhou]{li2026s2grag}
Minghan Li, Junjie Zou, Xinxuan Lv, Chao Zhang, and Guodong Zhou.
\newblock {S2G-RAG}: Structured sufficiency and gap judging for iterative
  retrieval-augmented {QA}.
\newblock In \emph{Proceedings of the 64th Annual Meeting of the Association
  for Computational Linguistics (Volume 1: Long Papers)}, pages 25846--25862,
  2026.
\newblock \doi{10.18653/v1/2026.acl-long.1185}.
\newblock URL \url{https://aclanthology.org/2026.acl-long.1185/}.

\bibitem[Kieback et~al.(2026)Kieback, Amadasun, Chadha, and
  Elkins]{kieback2026tasr}
Adrian Kieback, Uyiosa~Philip Amadasun, Aman Chadha, and Aaron Elkins.
\newblock {TASR}: Training-free adaptive stopping for iterative retrieval.
\newblock \emph{arXiv preprint arXiv:2606.13814}, 2026.
\newblock \doi{10.48550/arXiv.2606.13814}.
\newblock URL \url{https://arxiv.org/abs/2606.13814}.

\bibitem[Dong et~al.(2025)Dong, Jin, Li, Zhu, Dou, and Wen]{dong2025ragcritic}
Guanting Dong, Jiajie Jin, Xiaoxi Li, Yutao Zhu, Zhicheng Dou, and Ji-Rong Wen.
\newblock {RAG-Critic}: Leveraging automated critic-guided agentic workflow for
  retrieval augmented generation.
\newblock In \emph{Proceedings of the 63rd Annual Meeting of the Association
  for Computational Linguistics (Volume 1: Long Papers)}, pages 3551--3578,
  2025.
\newblock \doi{10.18653/v1/2025.acl-long.179}.
\newblock URL \url{https://aclanthology.org/2025.acl-long.179/}.

\bibitem[Geifman and El-Yaniv(2017)]{geifman2017selective}
Yonatan Geifman and Ran El-Yaniv.
\newblock Selective classification for deep neural networks.
\newblock In \emph{Advances in Neural Information Processing Systems}, 2017.
\newblock URL \url{https://arxiv.org/abs/1705.08500}.

\bibitem[Kang et~al.(2024)Kang, G{\"u}rel, Yu, Song, and Li]{kang2024crag}
Mintong Kang, Nezihe~Merve G{\"u}rel, Ning Yu, Dawn Song, and Bo~Li.
\newblock {C-RAG}: Certified generation risks for retrieval-augmented language
  models.
\newblock In \emph{Proceedings of the 41st International Conference on Machine
  Learning}, volume 235 of \emph{Proceedings of Machine Learning Research},
  2024.
\newblock URL \url{https://proceedings.mlr.press/v235/kang24a.html}.

\bibitem[Yang et~al.(2018)Yang, Qi, Zhang, Bengio, Cohen, Salakhutdinov, and
  Manning]{yang2018hotpotqa}
Zhilin Yang, Peng Qi, Saizheng Zhang, Yoshua Bengio, William Cohen, Ruslan
  Salakhutdinov, and Christopher~D. Manning.
\newblock {HotpotQA}: A dataset for diverse, explainable multi-hop question
  answering.
\newblock In \emph{Proceedings of the 2018 Conference on Empirical Methods in
  Natural Language Processing}, pages 2369--2380, 2018.
\newblock \doi{10.18653/v1/D18-1259}.
\newblock URL \url{https://aclanthology.org/D18-1259/}.

\end{thebibliography}

\appendix
\section{Descriptive Full-Set Table}

\begin{table}[H]
\centering
\small
\caption{Descriptive results on indices 0--999. These values are not the confirmatory analysis.}
\begin{tabular}{lrrrrr}
\toprule
System & EM & Searches & STOP P & STOP R & STOP AP \\
\midrule
Native Search-R1 & 0.4490 & 2.6100 & -- & -- & -- \\
Structured Base & 0.4490 & 2.6090 & 0.0000 & 0.0000 & 0.3661 \\
Old S2G LoRA & 0.4490 & 2.6090 & 1.0000 & 0.0015 & 0.3694 \\
Expanded S2G LoRA & 0.4430 & 2.5210 & 0.6163 & 0.0805 & 0.3987 \\
\bottomrule
\end{tabular}
\end{table}

\section{Frozen Claim Boundaries}

\begin{itemize}[leftmargin=*]
  \item Supported: the expanded S2G judge trained for Search-R1 reduces retrieval calls on the confirmatory test set while keeping the loss in answer accuracy within the two-percentage-point range specified before evaluation.
  \item Unsupported: the policy improves answer quality, certifies low STOP risk, dominates all fixed or matched-cost baselines, or lowers total inference cost.
  \item Scope: S2G-code-aligned HotpotQA distractor development indices 200--999, the frozen Search-R1 7B/E5/Wiki-2018/top-3/four-search configuration, and the frozen expanded adapter and threshold.
\end{itemize}

\end{document}